# Robust Real-time Segmentation of Bio-Morphological Features in Human Cherenkov Imaging during Radiotherapy via Deep Learning

Shiru Wang, Yao Chen, Lesley A. Jarvis, Yucheng Tang, David J. Gladstone, Kimberley S. Samkoe, Brian W. Pogue, Petr Bruza, and Rongxiao Zhang

*Abstract*—Cherenkov imaging enables real-time visualization of megavoltage X-ray or electron beam delivery to the patient during Radiation Therapy (RT). Bio-morphological features, such as vasculature, seen in these images are patient-specific signatures that can be used for verification of positioning and motion management that are essential to precise RT treatment. However until now, no concerted analysis of this biological feature-based tracking was utilized because of the slow speed and accuracy of conventional image processing for feature segmentation. This study demonstrated the first deep learning framework for such an application, achieving video frame rate processing. To address the challenge of limited annotation of these features in Cherenkov images, a transfer learning strategy was applied. A fundus photography dataset including 20,529 patch retina images with ground-truth vessel annotation was used to pre-train a ResNet segmentation framework. Subsequently, a small Cherenkov dataset (1,483 images from 212 treatment fractions of 19 breast cancer patients) with known annotated vasculature masks was used to fine-tune the model for accurate segmentation prediction. This deep learning framework achieved consistent and rapid segmentation of Cherenkov-imaged bio-morphological features on another 19 patients, including subcutaneous veins, scars, and pigmented skin. Average segmentation by the model achieved Dice score of 0.85 and required less than 0.7 milliseconds processing time per instance. The model demonstrated outstanding consistency against input image variances and speed compared to conventional manual segmentation methods, laying the foundation for online segmentation in real-time monitoring in a prospective setting.

## I. INTRODUCTION

RADIATION therapy (RT) has been established as the standard of care for many cancers, either stand alone or adjuvant with surgery and chemotherapy, significantly improving survival outcomes [1]. For breast cancer, as the most common malignancy in women, RT is commonly used to eliminate residual cancer cells post-surgery therapy reducing the risk of recurrence in advanced stages or as neoadjuvant therapy to shrink tumors in locally advanced cases before surgery to make them easier to remove.

For general RT procedures and breast irradiation specifically, it is crucial to accurately target the tumor while sparing surrounding healthy tissue [2]. Excessive deviations in treatment delivery and patient setup can lead to negative consequences including suboptimal treatment to the tumor, elevated healthy tissue toxicity and secondary radiation induced cancer risks [3]. Optical Surface Guided RT (SGRT) became state-of-the-art for precise patient setup verification and monitoring [4], by allowing the patient's outer anatomy to be accurately aligned with the treatment machine prior to each treatment. SGRT relies exclusively on surface profile and hence is not sensitive to sub-surface changes. To mitigate such limitations, SGRT is typically utilized in combination with other onboard imaging guidance such as x-ray planar imaging and/or cone beam CT.

Adding to conventional SGRT techniques, Cherenkov imaging has emerged as a new SGRT component for real-time visualization of RT, as depicted in Fig. 1(a) [5]. Cherenkov light emission occurs when high-energy photon or electron beams interact with tissue, inducing visible light emission that resembles the radiation beam shape and superficial dose on the patient surface, as shown in Fig. 1(b) [6], [7]. The current implementation of this technique utilizes intensified cameras that are time-gated to synchronize with the linear accelerator pulses, enabling the

Manuscript submitted on September 9, 2024. This work was supported by the U.S. National Institutes of Health under Grants R01 EB023909-08, R44 CA265654-02, and P30 CA023108-41. *(Shiru Wang and Yao Chen are co-first authors. Corresponding author: Yao Chen; Rongxiao Zhang.)*

S. Wang, Y. Chen, D. J. Gladstone, K. S. Samkoe, and P. Bruza are with Thayer School of Engineering, Dartmouth College, Hanover, NH 03755 USA (e-mail: shiru.wang.th@dartmouth.edu; yao.chen.th@dartmouth.edu; david.gladstone@dartmouth.edu; kimberley.s.samkoe@dartmouth.edu; petr.bruza@dartmouth.edu).
L. A. Jarvis is with Department of Radiation Oncology, Dartmouth Health, Lebanon, NH 03756 USA (e-mail: lesley.a.jarvis@hitchcock.org).
Y. Tang is with NVIDIA Corporation, Santa Clara, CA 95051 USA (e-mail: yuchengt@nvidia.com).
B. W. Pogue is with University of Wisconsin-Madison, Madison, WI 53705 USA (e-mail: bpogue@wisc.edu).
R. Zhang is with University of Missouri-Columbia, Columbia, MO 65211 USA (e-mail: rzkp2@health.missouri.edu).

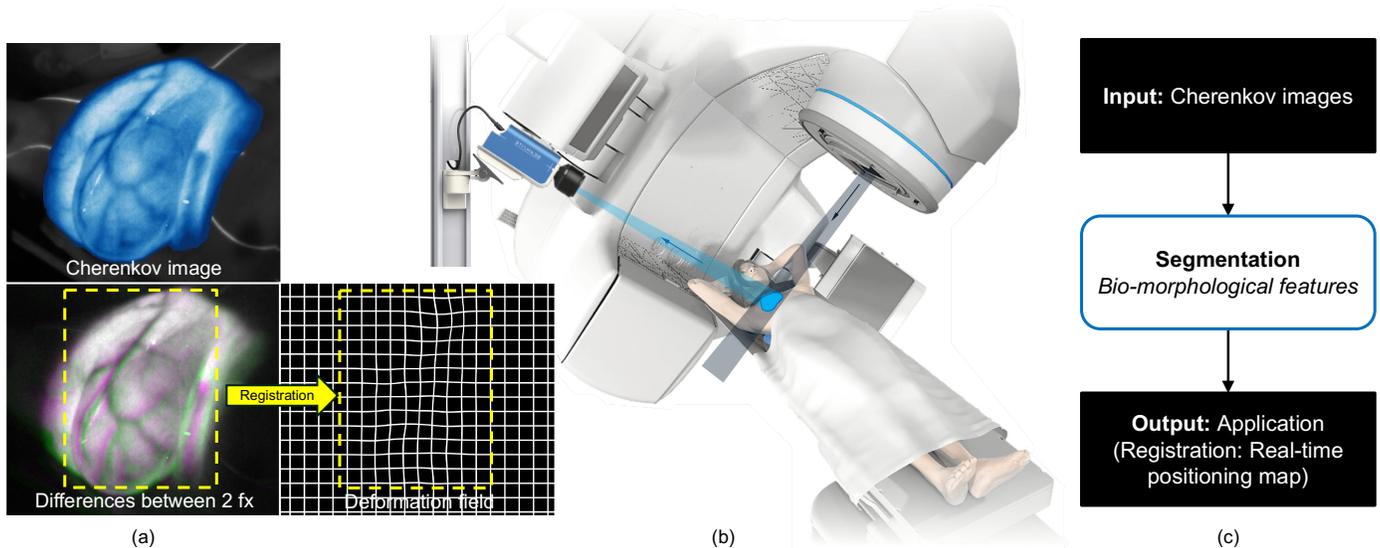

**Fig. 1.** Cherenkov imaging during radiotherapy. (a) Cherenkov image from breast RT. The top panel shows an example Cherenkov image of a breast cancer patient undergoing radiotherapy, with a pseudo-colormap for enhanced visualization. The bottom left panel displays an overlay of Cherenkov images from two-day treatment fractions exhibiting the setup variations between them, where magenta represents the variation in the later fraction and green represents the residual in the former fraction. The bottom right panel displays the deformation between the two fractions (fx) that is quantified by registration of bio-morphological features in the yellow box region of Cherenkov images. (b) Setup of Cherenkov imaging system during RT, providing real-time Cherenkov imaging. (c) Proposed online workflow based on the segmentation of bio-morphological features from Cherenkov imaging for real-time monitoring and verification. The focus of this work is the segmentation of bio-morphological features shown in the middle step of the workflow, enabling the ultimate monitoring and validating the patient positioning in real time.

real-time capture of low-intensity Cherenkov light while effectively suppressing ambient room light [8], [9]. This technology works synergistically with traditional SGRT and offers novel applications in RT, including geometric treatment delivery verification and dose monitoring [10], [11].

Cherenkov images acquired from RT patients displays very noticeable patient-specific bio-morphological features that originate from strong light absorption of sub-surface vasculature [12], [13], observed in Fig. 1(a). These bio-morphological features could be used as an intrinsic signature of each individual patient, to verify their positioning, or to quantify locoregional deformation, as shown in Fig. 1(c), that have been neglected or underestimated. The feasibility has been demonstrated in a previous study but the translation to real-time application was limited by the speed and consistency of manual segmentations. A robust and rapid segmentation of the bio-morphological features from Cherenkov imagery is key to using this in online positioning verification and real-time monitoring.

Deep learning (DL)-based neural networks have been extensively tested for performing image segmentation tasks across various modalities of medical imaging data, including MRI, CT, PET, and other optical imaging techniques [14] [15]. Multiple neural network structures have been developed for performing image segmentation tasks. UNet is a convolutional neural network (CNN) architecture designed for biomedical image segmentation [16], known for its U-shaped structure that facilitates image spatial information mining and features extraction for semantic segmentation. ResNet, or Residual Network [17], is a deep learning model that utilizes residual connections to enable the training of very deep networks, enhancing performance in image segmentation tasks by mitigating the vanishing gradient problem. Both deep learning architectures are extensively employed in the medical image segmentation area and have realized satisfactory results. In addition to classic networks, recent models like Segment Anything Model (SAM) [18] and SAM2 [19] which trained with billions of annotated images have also demonstrated generalizable performances in image segmentation tasks.

In this study, we applied ResNet algorithm to perform robust and automatic segmentation of bio-morphological features that appeared in Cherenkov imaging. A transfer learning strategy was employed by leveraging pre-trained models on a large dataset of fundus photographs, which include retinal vasculature with similar morphology to subcutaneous veins observed in Cherenkov images. This first-time application of a deep learning algorithm based on transfer learning in Cherenkov imaging reliably extracts patient-specific bio-morphological features rapidly, paving the way for similar applications in a real-time and prospective setting.

## II. METHODOLOGY

### A. Problem Formulation

The existing Cherenkov image data captured during human RT is limited since the technology has only recently been available to clinics. Additionally, Cherenkov images often suffer from low resolution, making it difficult for clinicians to discern various bio-morphological features. As a result, the key problem in developing a machine learning based segmentation algorithm is the lack of sufficiently large and annotated datasets to train the DL model.

To address this problem, a transfer learning task was formulated[20]. Given a source domain $D_s$ and a related source task $T_s$, a corresponding target domain $D_T$ and a target task $T_T$ with insufficient labeled data $n_T$, the objective is to improve the learning of the predictive function $f_T(\cdot)$ in $D_T$ by leveraging the knowledge gained from the $D_s$ and $T_s$. Here, $D_s$ is the retina vessel image and $D_T$ is the clinical Cherenkov image. $T_s$ and $T_T$ are all binary segmentation tasks. With $\mathcal{L}_T$ as the expected loss over the target domain, the first pre-train step can be formulated as:

$$\min_{f_T} \mathcal{L}_T = \mathbb{E}_{(\mathbf{x},y) \sim \mathcal{D}_T}[\ell(f_T(\mathbf{x}), y)] \quad (1)$$

where $\ell(\cdot)$ is the loss function, x is the input image and y is the corresponding ground truth binary mask It uses a model $f_s(\cdot)$ trained on the source task:

$$f_T(\mathbf{x}) = f_S(\mathbf{x}) + \Delta f(\mathbf{x}) \quad (2)$$

where $\Delta f(\mathbf{x})$ represents the adaptation of the source model $f_s(\cdot)$ to $T_T$.

After obtaining a set of weights with large amount of training data in $D_s$, the next step is fine-tuning the model $f_s(\cdot)$ on the small dataset from $D_T$, such that the model can adapt to $T_T$ by minimizing:

$$\min_{\Theta_T} \mathbb{E}_{(\mathbf{x},y) \sim \mathcal{D}_T}[\ell(f_T(\mathbf{x}; \Theta_T), y)] \quad (3)$$

where $\Theta_T$ denotes the parameters of the model after fine-tuning.

### B. Cherenkov Imaging Transfer Learning

---

**Algorithm 1** Transfer Learning

**Inputs**: $D_{source}$, $D_{target}$, step size η, transfer rate
**Outputs**: Fine-tuned model f
Initialize the model parameters on the source domain;
**while** not converged **do**
compute combined Dice and CE loss on $D_{source}$, $D_{target}$;
update model parameters using η and λ;
**end**
**return** f

---

In the first step of the transfer learning strategy, aiming to solve the primary challenge mentioned in section II.A, the dataset used for training was the color fundus image vessel segmentation (FIVES) dataset, which contains 800 high-resolution multi-disease color fundus photographs with pixelwise manual annotation [21]. In order to increase the training sample size to ensure the model's generalizability, we used a patch-wise training strategy. Each original retina image and its corresponding ground-truth (GT) in FIVES dataset was cropped into 36 patches, each with size of 224 pixels by 224 pixels, in the center regions of interest (ROIs) as shown in Fig. 2(a).

To improve model generalizability, a data augmentation strategy was performed by applying the following transformations to training dataset, horizontal flip for 33% of data, random rotation with one of the angles (90, 180, or 270 degrees) for 33% of data, adding random Gaussian noise for 50% of data with maximum magnitude of 25%.

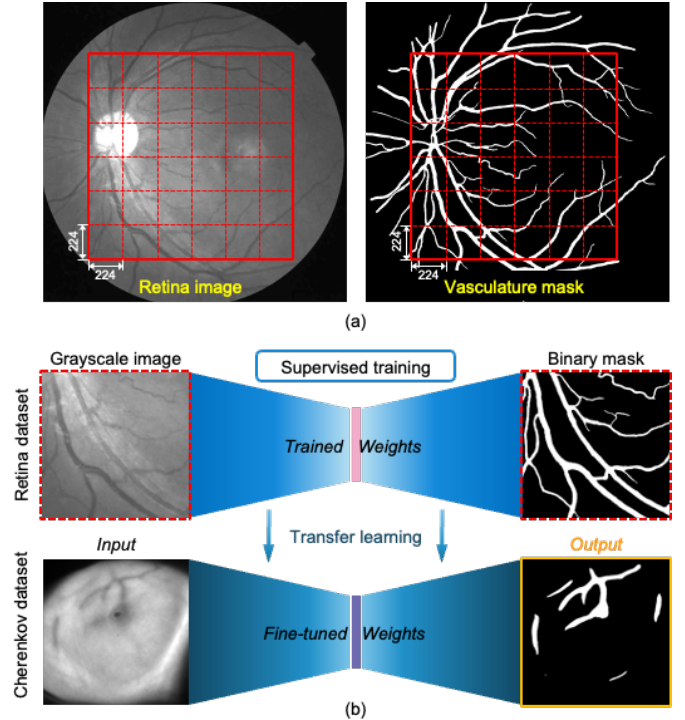

Fig. 2. Model training methodology. (a) Retina photograph example in FIVES dataset with corresponding vasculature mask. In the training stage, each retina image was cropped into 36 square patches with size of 224 pixels by 224 pixels or data augmentation. (b) Paradigm of transfer learning. The deep learning model is first trained in a supervised manner using the FIVES retinal dataset. Then pre-trained weights were fine-tuned and used for segmentation of bio-morphological features in Cherenkov images.

To make the training more efficient, we retained only patches where the labeled area exceeded 5% of the total patch area. In total, 20,529 retina patches were used for training. The patches were randomly split into 90% for training and 10% for validation.

The bio-morphological characteristics of these images from FIVES dataset closely resemble those of Cherenkov vessel images. It is a good source for pre-training our model to learn general features such as vessel shapes and boundary conditions. As shown in Fig. 2(b), after getting a set of pre-trained weights, Cherenkov images were fed into the same model for fine-tuning. We picked the weights from the smallest validation loss in the fine-tuning round as the final weights for our model. The aim for this was to let the model learn more specific features in the Cherenkov vessel images.

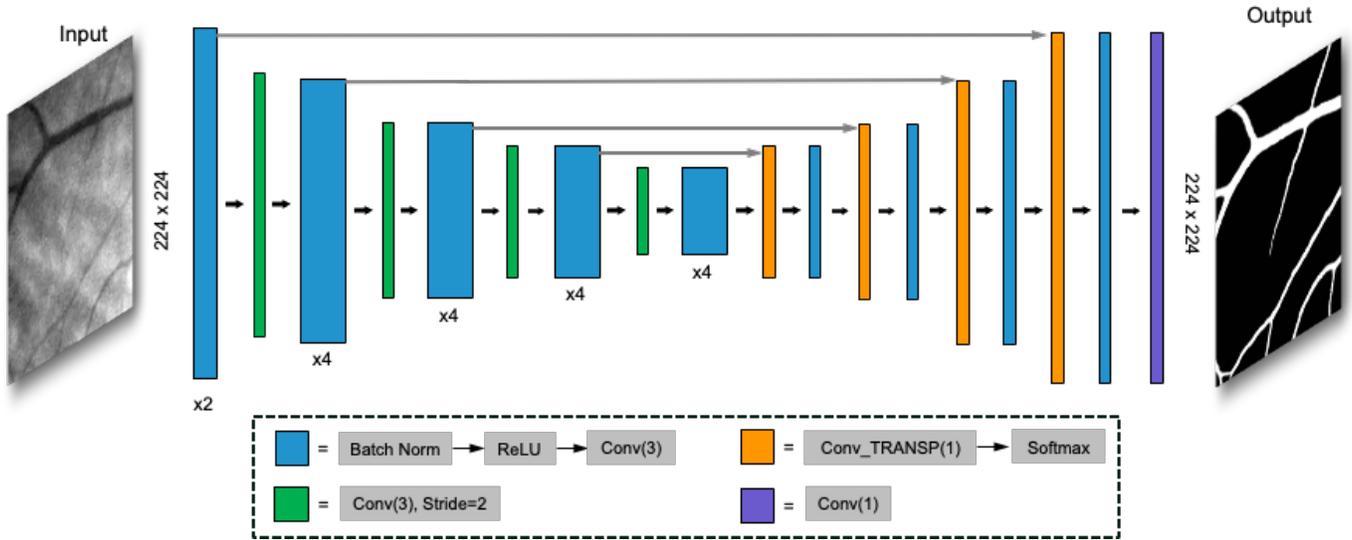

Fig. 3. Architecture of the SegResNet2D. The input data are N by 224 by 224 retina or Cherenkov grayscale images, where N is the total number of training dataset. The outputs are binary masks of segmented vasculature with the same size of the input images. Detailed steps involved in each layer of architecture are described in the dashed line box below. 'x2' and 'x4' indicate repeating the described steps by looping through from the first to the last step, two or four times, respectively.

### C. Segmentation Network

SegResNet 2D network with MONAI [18] were implemented in this study as the segmentation backbone as shown in Fig. 3. The core for encoder part is the usage of ResNet blocks [17]. In each block, a convolutional network (CNN) with kernel size of 3 was used to learn the image features. The original number of filters was equal to 32. Batch normalization was applied before CNN approach to make the training process faster and more stable. Each block's output was followed by an additive identity skip connection, with the goal of eliminating the gradient vanishing problem [17]. The decoder structure was symmetric to the encoder part. However, only a single block including one convolutional network and a *softmax* layer was used in each convolutional level. A sigmoid function was applied to the output layer of this network. The end of the decoder was the segmented mask which had the same spatial size as the original input images.

### D. Overall Objectives

The SegResNet [17] was trained by jointly optimizing a Dice loss function and a Cross Entropy (CE) loss. This enables the model to learn general visual representations as well as variances among all segmentation tasks. A sigmoid function was applied to the output in each training epoch before calculating the loss. The overall objectives can be formulated as:

$$\mathcal{L}_{seg} = \lambda_1 * \mathcal{L}_{Dice}^w(y_t, y_p) + \lambda_2 * \mathcal{L}_{CE}^w(y_t, y_p) \quad (4)$$

Where $y_t$ represents the ground-truth mask and $y_p$ represents the predicted mask. $\lambda_1$ and $\lambda_2$ are hyperparameters that balance the loss functions. In our practical implementation, the training effect is best when we set $\lambda_1 = 1$ and $\lambda_2 = 0.1$.

### E. Model Training Hyperparameters

Each model was implemented using *PyTorch* on an Nvidia GeForce RTX 2080Ti GPU. A batch size of 24 and Root Mean Square Propagation (*RMSProp*) optimizer with a learning rate of 10-5 were used for all models. For pre-training models, 400 epochs were applied using a 10-8 L2 regularization decay rate. For the fine-tune task, we trained the model for 100 epochs and picked the best weights in the validation round as the final model weights. The training curves of performance are displayed in the Fig. 1 of Supplementary File 1.

### F. Evaluation Metrics

For the segmentation tasks, Dice score, intersection over-union (IoU) score, and boundary IoU score were used to evaluate the accuracy of the results.

## III. EXPERIMENTS AND EVALUATION

### A. Cherenkov Imaging Dataset

A dataset of Cherenkov images of breast cancer patient during RT was used for model fine tuning and testing. The patient images involved in this study were collected from a retrospective study approved by the Institutional Review Board (IRB) of Dartmouth Health. Two Cherenkov cameras (BeamSite, DoseOptics LLC, Lebanon NH, USA) were mounted on the ceiling of the treatment room at Dartmouth Hitchcock Medical Center and recorded Cherenkov emission at 19.6 frames-per-second (fps) during RT beam delivery. Cherenkov video frames were aggregated into a cumulative image for each entire treatment fraction. Both the raw video footage and the cumulative image were preserved for subsequent analysis.

In total, 385 Cherenkov images from 36 breast cancer patients and 2 osteosarcoma patients (38 patients in total) treated with RT were included in this study. 212 Cherenkov images from 19 breast patients were used for model fine tune and the rest of 171 Cherenkov images from 17 breast cancer patients and 2 osteosarcoma patients (19 patients in total) were used for model testing. Experienced observers (Y.C., L.A.J.) manually annotated masks of subcutaneous veins of patients' breast surface in Cherenkov images as GT for fine-tuning the model. Detailed information of the clinical Cherenkov dataset is summarized in Table 1.

### TABLE 1

PATIENT DEMOGRAPHICS IN CHERENKOV IMAGING DATASET FOR MODEL FINE TUNE AND TEST. (M – MALE, F – FEMALE, LT – LEFT BREAST, RT – RIGHT BREAST, IMGS - IMAGES).

| Patient data for model fine tune | | | | Patient data for model test | | | |
|---|---|---|---|---|---|---|---|
| No. | Age/Sex | Site | Imgs | No. | Age/Sex | Site | Imgs |
| 1 | 63/F | Lt | 6 | 20 | 54/F | Rt | 13 |
| 2 | 69/F | Lt | 8 | 21 | 53/F | Lt | 14 |
| 3 | 69/F | Lt | 7 | 22 | 60/F | Rt | 5 |
| 4 | 51/F | Lt | 8 | 23 | 52/F | Lt | 16 |
| 5 | 68/F | Lt | 7 | 24 | 69/F | Rt | 7 |
| 6 | 58/F | Rt | 11 | 25 | 64/F | Lt | 13 |
| 7 | 67/F | Rt | 10 | 26 | 50/F | Rt | 3 |
| 8 | 66/F | Rt | 5 | 27 | 55/F | Lt | 8 |
| 9 | 70/F | Rt | 3 | 28 | 54/F | Rt | 12 |
| 10 | 73/F | Rt | 8 | 29 | 57/F | Lt | 14 |
| 11 | 71/F | Rt | 6 | 30 | 60/F | Lt | 14 |
| 12 | 71/F | Rt | 6 | 31 | 76/F | Lt | 14 |
| 13 | 65/F | Lt | 24 | 32 | 79/F | Rt | 14 |
| 14 | 76/F | Rt | 9 | 33 | 51/F | Lt | 4 |
| 15 | 77/F | Rt | 18 | 34 | 70/F | Lt | 16 |
| 16 | 74/F | Lt | 20 | 35 | 73/F | Lt | 1 |
| 17 | 77/F | Rt | 22 | 36 | 59/F | Lt | 1 |
| 18 | 52/F | Lt | 19 | 37 | 55/M | Leg | 1 |
| 19 | 54/F | Lt | 15 | 38 | 70/M | Leg | 1 |

For the preparation of Cherenkov image data, the beam regions were selected as ROIs and cropped to match a size that is an exact multiple of the patch size used for the retina images (N * 224, where N is an integer, N ≤ 4). We first selected the ROIs based on vessel locations in the Cherenkov images and then decided on the number of patches which could be cut in each image. The number of patches per Cherenkov image varied from 4 to 16. In total, 1,483 Cherenkov vessel patches were used for finetuning the pre-trained model. We followed the same training/validation protocols, i.e., the patches were randomly split into 90% for training and 10% for validation.

### B. Model Evaluation

#### 1) Prediction Robustness Against Variations

In clinical practice, Cherenkov images often exhibit a low signal-to-noise ratio due to scatter radiation noise, and the presence of background light and thus increased shot noise is present after background subtraction. To test the predication consistency of our model under different cases, we first applied it to original Cherenkov images to generate segmentation masks, which were severed as references. Several operations including rotation and adding random Gaussian noise were then applied to the original Cherenkov images in order to mimic the real case scenario. The same model was then tested on modified images and compared with the references. Dice scores were used here as the evaluation metrics.

#### 2) Prediction Consistency and Speed Compared to Manual Segmentation

In previous studies, manual segmentation was adopted to delineate bio-morphological features. This process is slow and is associated with user variance. A comparative study was conducted against the manual segmentation.

Five experienced individuals who were familiar with vessel identification in two Cherenkov images were selected. Each of the five participants independently segmented both images three times over three different days. The same two images were also tested on our model three times. Consistency across three days per individual, as well as between human and model predictions, was evaluated for both cases using the Dice score and boundary IoU.

#### 3) Segmentation on Sub-Cumulative Frames from Raw Cherenkov Videos

The Cherenkov video stream is captured and displayed at a frame rate of 19.6 fps. For real time online applications, it is imperative to perform analysis on sub-cumulative images corresponding to a range of temporal resolution with typically increased noise and possible motion interplays. The performance on such sub-cumulative images was quantified utilizing data acquired with and without respiratory motion. A free breathing (i.e., with respiratory motion) and a deep inspiration breath hold (DIBH) (i.e. no respiratory motion) patients' breast Cherenkov video streams were used as the test data. For each patient, we first summed all frames to achieve a cumulative Cherenkov image and used the well-trained model to generate a segmented mask which served as the reference. Then different time-gates ranging from 10 to 120 frames, corresponding to cumulations from 0.5 sec to 6.1 sec, were selected and summed to several sub-cumulative Cherenkov images. The same model was used to segment all sub-cumulative images, and the result masks were compared with the reference separately. Dice scores were used to evaluate the performance of the model.

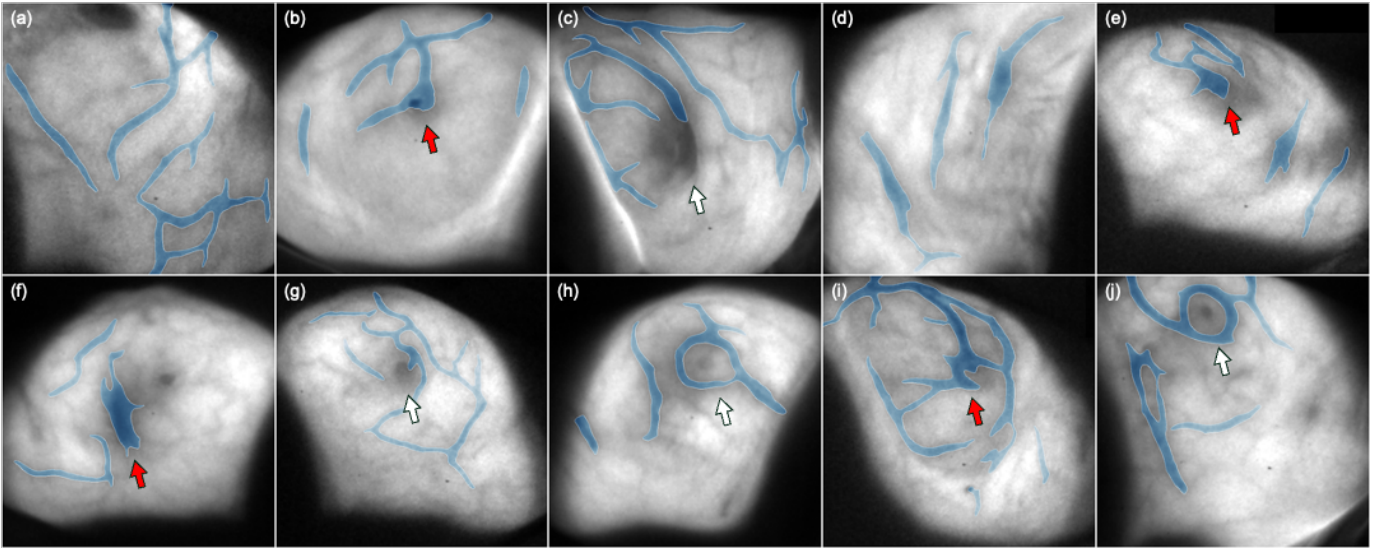

Fig. 4. Segmentation visualization of Cherenkov-imaged bio-morphological features on 10 representative breast cancer patients. (a) - (j) Segmented bio-morphological features by fine-tuned SegResNet model with enhanced edges transparently overlaid on Cherenkov images for 10 breast patients. Segmented bio-morphological features are mainly subcutaneous veins in the breast surface but partly include other features like scars and nipples. Red arrows in panel (b), (e), (f), and (i) indicate false inclusion of segmentation with nipple and scars, while white arrows in panel (c), (g), (h), and (j) indicate the well-performed predictions on accurately exclusively identifying vasculature.

## IV. RESULTS

The segmentation visualization of bio-morphological features in Cherenkov images of 10 representative breast patients are shown in Fig. 4. In addition, we assessed the performances of speed and robustness of the proposed DL segmentation framework in three different experiments compared with manually segmentation method. Both quantitative and qualitative results are introduced in this section.

### A. Model Prediction Robustness Against Variations

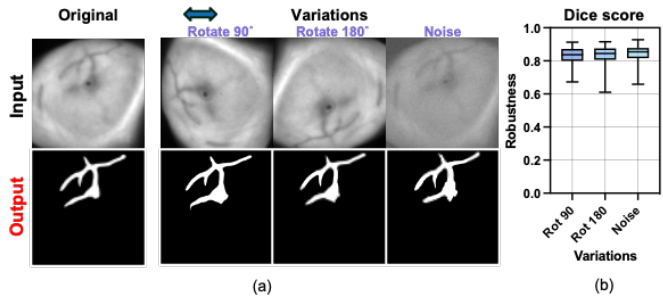

Fig. 5. Assessment of model robustness. (a) Qualitative segmentation results of an example Cherenkov image with variation of performing rotation and adding noise. (b) Similarity between original prediction and predictions with variations among all Cherenkov testing images.

Visualization segmentation results are shown in Fig. 5(a) aligned with the input images. The largest connected vessel area was used in the example to compare with the reference. Despite some changes in detailed features, the overall shapes closely matched the reference. We then applied the same method to our entire test dataset and calculated the Dice scores, as shown in Fig. 5(b). Dice scores of $0.83\pm0.05$, $0.84\pm0.05$, and $0.85\pm0.04$ were found for variations of rotation

90 degrees, 180 degrees, and adding random noise with magnitude of 10% among all testing Cherenkov images. High Dice scores indicate that our proposed model has good generalizability to the rotation and noise, so it could be used under different conditions.

### B. Model Prediction Consistency and Speed

Two Cherenkov images were provided, as shown in Fig. 6(a). The top and bottom image represent a simple case and more complex case respectively. The quantitative comparison results, shown in Fig. 6(b), indicated that our proposed model had better consistency in segmentation across different days compared to the human participants.

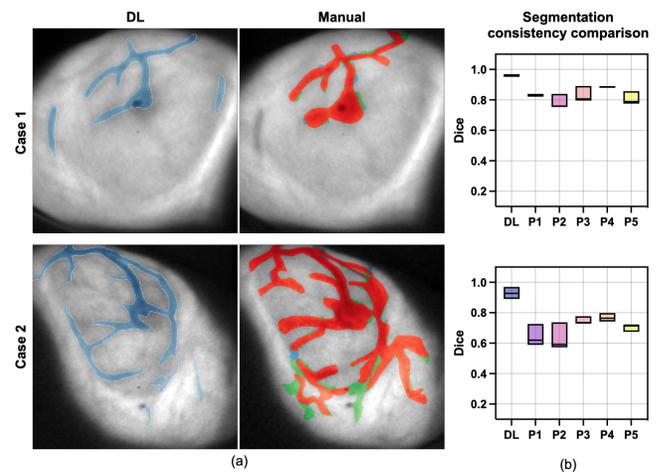

Fig. 6. Deep learning versus manual segmentation. (a) Visualization of prediction made by DL and three repeated (red, green, blue respectively represents three repetition) manual segmentation by one experienced observer on two Cherenkov cases. (b) Segmentation consistency among three repetitions by DL compared to 5 observers on two cases respectively.

Moreover, manual segmentation showed poorer performance with the relatively more complex case. The differences in segmentation results increased in case two compared to case one. In contrast, our model maintained consistent performance, even as the task's complexity increased.

The segmentation time for the deep learning method takes about 0.7 milliseconds per case, regardless of the complexity of bio-morphological features, whereas manual segmentation required 2 to 3 minutes, marking a conservatively estimated $\times 10^6$ improvement in terms of speed.

## C. Model Segmentation on Sub-Cumulative Frames from Raw Cherenkov Videos

In Fig. 7(a), the same model was used to segment all sub-cumulative frame images and compare with the reference (cumulative image from entire video). When the number of cumulative frames increased, the Dice score also increased, which was expected since the SNR increased. Moreover, our model showed good consistency on sub-cumulative images with a Dice score over 0.8 on only about 35 frames. This showed the potential application of our proposed model for real-time online Cherenkov image analysis.

Similarly, in Fig. 7(b), Cherenkov video stream of a free-breathing patient was tested following the same steps. Dice score increased at first and then decreased which shows different trends to the DIBH patient. A consistent increasing trend in segmentation similarity was observed in the DIBH patient with increasing cumulative frame numbers, where a small elevation was observed in FB patient. This is because of respiratory motion. When patients breathe, the positions of their breasts are changed accordingly. During the post-analysis processing, we summarized video frames with motion artifacts, The movements of the bio-morphological features like vessel positions led to Dice score variations since the model is sensitive to changes of the input images.

## V. Discussion

Transfer learning, a technique where a pre-trained model is adapted to a new but related task, brings significant advantages in prediction performance, especially in areas where annotated data is limited for a reliable supervised training [22]. The pre-trained model included complicated feature detectors that have been optimized on extensive data, aiding in potentially higher accuracy when applied to the new task. In this specific work of segmenting Cherenkov-imaged bio-morphological features, the usage of transfer learning dramatically reduced the need for extensive labeling of Cherenkov data. This approach allows for the efficient adaptation of sophisticated DL model to this niche imaging modality, significantly enhancing segmentation accuracy and robustness. Furthermore, by using the learned features from other similar imaging data, transfer learning expedites the training process, leading to faster deployment in segmenting bio-morphological features from Cherenkov imagery.

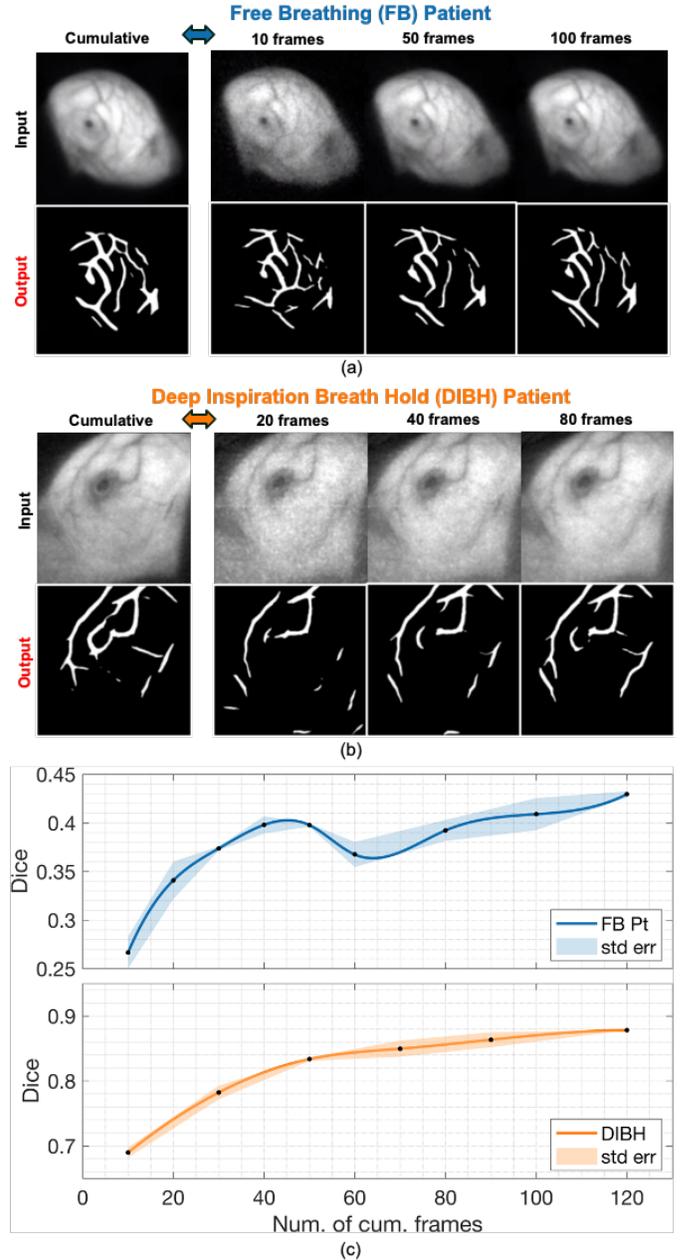

**Fig. 7.** Model segmentation performance on raw Cherenkov video frames with different numbers of cumulative frames. Each sub-cumulative frame segmentation was compared to the whole time-series cumulation segmentation using a Dice score. (a) Segmentation on a free breathing (FB) patient. (b) Segmentation on a deep inspiration breath hold (DIBH) patient. (c) The blue curve at the top represents the FB patient, while the orange curve at the bottom represents the DIBH patient. The shaded areas in the plot represent the standard errors of temporal sub-cumulative frames.

In addition, zero-shot segmentation is a popular method aimed at improving model performance with limited labeled data. This approach seeks to recognize new classes without prior knowledge by leveraging semantic relationships between known and unknown classes. However, when applied to Cherenkov image segmentation, models like SAM and SAM2 struggled with accuracy and robustness, particularly on detailed spatial bio-morphological features—two critical aspects for this task. These challenges may stem from image noise, contrast variances, and domain differences from the data on which these large models were trained. In contrast, the proposed model, utilizing transfer learning strategies, demonstrated superior performance on Cherenkov datasets, making it more suitable for clinical use.

This DL method's efficiency and speed are critical, especially for online applications involving real-time monitoring of patient positioning. In scenarios like evaluating patient positioning accuracy during RT, the ability to process data swiftly and consistently in terms of bio-morphological features identification can really make the difference between timely action and harmful delays. The method's rapid processing capabilities ensure that data can be analyzed as it streams, maintaining a high throughput necessary for real-time applications. Furthermore, consistency in output, achieved through robust model training and validation, ensures that the systems' performance remains stable under various conditions, boosting its dependability. Therefore, by optimizing both speed and consistency, the proposed method can meet the demands of real-time monitoring systems, providing timely and reliable output.

Currently, 212 Cherenkov images with annotated labels from 19 breast patients were used for model fine-tuning. Moving forward, more retrospective Cherenkov data in breast and other sites will be collected. Larger datasets can help in training more robust models, reducing the variance and improving the overall generalization to unseen data like more subtle patterns within Cherenkov data. Continuous data acquisition opens avenues for longitudinal studies and model updates, enabling the system to expand to other treatment sites such as abdomen and extremity over time and maintaining high accuracy in its predictions.

The ongoing efforts to integrate the segmentation capabilities developed in this work into a DL registration algorithm are pivotal for enhancing quantitative and real-time monitoring and clinical quality control of patient positioning. Segmentation provides a detailed understanding of the data, which can then be used to align latter images more accurately with reference images. This combined approach would facilitate more precise and faster tracking of changes and uncertainties over time. This integration would enable the monitoring and early detection of significant events or detrimental deviations. We propose to use VoxelMorph [23], a fast unsupervised learning framework for deformable image registration, to rapidly calculate the deformation field between a pair of Cherenkov-imaged bio-morphological features. Through this synergy of segmentation and registration, real-time systems can achieve timely display of patient positioning verification, thereby supporting faster decision-making during dose delivery and resulting in higher precision and more informed clinical practice.

## VI. Conclusion

For the first time, the SegResNet model for bio-morphological feature segmentation has been used successfully in extraction of bio-morphological information within clinical Cherenkov images. The usage of this transfer learning strategy efficiently addressed the scarcity of annotated Cherenkov imaging data. By training the model on large datasets with similar bio-morphological features and fine-tuning it with clinical Cherenkov images, we accelerated training times and improved accuracy, enabling the model to generalize effectively with limited data. The well-trained model was able to process a single Cherenkov image in approximately 0.7 milliseconds, which is approximately $10^6$ faster than traditional segmentation methods. Moreover, the model maintained a Dice score exceeding 0.85, demonstrating robustness against noise and varying rotation angles, thus ensuring reliable performance in dynamic clinical environments. These two aspects are essential for real-time online application of patient position setup during RT, where precise positioning and timely intervention is critical.

### Acknowledgment

The authors are grateful to Dr. Savannah M. Decker and Wilson Schreiber for providing assistance on collecting clinical Cherenkov data; and to Jacob P. Sunnerberg, Caleb Y. Kwon, Austin M. Sloop for their time and effort to provide input on manual segmentation of bio-morphological features.

### References

[1] R. Baskar, K. A. Lee, R. Yeo, and K.-W. Yeoh, "Cancer and Radiation Therapy: Current Advances and Future Directions," *Int J Med Sci*, vol. 9, no. 3, pp. 193–199, Feb. 2012, doi: 10.7150/ijms.3635.

[2] D. A. Jaffray, "Image-guided radiotherapy: from current concept to future perspectives," *Nat Rev Clin Oncol*, vol. 9, no. 12, pp. 688–699, Dec. 2012, doi: 10.1038/nrclinonc.2012.194.

[3] W. Bogdanich, "Radiation Offers New Cures, and Ways to Do Harm," *The New York Times*, Jan. 23, 2010. Accessed: Nov. 13, 2023. [Online]. Available: https://www.nytimes.com/2010/01/24/health/24radiation.html

[4] L. A. Jarvis *et al.*, "Initial Clinical Experience of Cherenkov Imaging in External Beam Radiation Therapy Identifies Opportunities to Improve Treatment Delivery," *International Journal of Radiation Oncology*Biology*Physics*, vol. 109, no. 5, pp. 1627–1637, Apr. 2021, doi: 10.1016/j.ijrobp.2020.11.013.

[5] L. A. Jarvis *et al.*, "Cherenkov Video Imaging Allows for the First Visualization of Radiation Therapy in Real Time," *International*

*Journal of Radiation Oncology*Biology*Physics*, vol. 89, no. 3, pp. 615–622, Jul. 2014, doi: 10.1016/j.ijrobp.2014.01.046.

[6] F. Newman, M. Asadi-Zeydabadi, V. D. Durairaj, M. Ding, K. Stuhr, and B. Kavanagh, "Visual sensations during megavoltage radiotherapy to the orbit attributable to Cherenkov radiation," *Med Phys*, vol. 35, no. 1, pp. 77–80, Jan. 2008, doi: 10.1118/1.2815358.

[7] K. D. Steidley, R. M. Eastman, and R. J. Stabile, "Observations of visual sensations produced by Cerenkov radiation from high-energy electrons," *Int J Radiat Oncol Biol Phys*, vol. 17, no. 3, pp. 685–690, Sep. 1989, doi: 10.1016/0360-3016(89)90125-9.

[8] D. A. Alexander *et al.*, "One Year of Clinic-Wide Cherenkov Imaging for Discovery of Quality Improvement Opportunities in Radiation Therapy," *Practical Radiation Oncology*, vol. 13, no. 1, pp. 71–81, Jan. 2023, doi: 10.1016/j.prro.2022.06.009.

[9] J. M. Andreozzi, R. Zhang, A. K. Glaser, L. A. Jarvis, B. W. Pogue, and D. J. Gladstone, "Camera selection for real-time in vivo radiation treatment verification systems using Cherenkov imaging," *Med Phys*, vol. 42, no. 2, pp. 994–1004, Feb. 2015, doi: 10.1118/1.4906249.

[10] S. M. Decker *et al.*, "Technical note: Visual, rapid, scintillation point dosimetry for in vivo MV photon beam radiotherapy treatments," *Med Phys*, Apr. 2024, doi: 10.1002/mp.17071.

[11] S. M. Decker *et al.*, "Performance comparison of quantitative metrics for analysis of in vivo Cherenkov imaging incident detection during radiotherapy," *BJR*, vol. 95, no. 1137, p. 20211346, Sep. 2022, doi: 10.1259/bjr.20211346.

[12] R. Zhang *et al.*, "Cherenkoscopy based patient positioning validation and movement tracking during post-lumpectomy whole breast radiation therapy," *Phys. Med. Biol.*, vol. 60, no. 1, p. L1, Dec. 2014, doi: 10.1088/0031-9155/60/1/L1.

[13] S. M. Decker *et al.*, "Estimation of diffuse Cherenkov optical emission from external beam radiation build-up in tissue," *JBO*, vol. 26, no. 9, p. 098003, Sep. 2021, doi: 10.1117/1.JBO.26.9.098003.

[14] A. Hosny, C. Parmar, J. Quackenbush, L. H. Schwartz, and H. J. W. L. Aerts, "Artificial intelligence in radiology," *Nat Rev Cancer*, vol. 18, no. 8, pp. 500–510, Aug. 2018, doi: 10.1038/s41568-018-0016-5.

[15] M. H. Hesamian, W. Jia, X. He, and P. Kennedy, "Deep Learning Techniques for Medical Image Segmentation: Achievements and Challenges," *J Digit Imaging*, vol. 32, no. 4, pp. 582–596, Aug. 2019, doi: 10.1007/s10278-019-00227-x.

[16] O. Ronneberger, P. Fischer, and T. Brox, "U-Net: Convolutional Networks for Biomedical Image Segmentation," in *Medical Image Computing and Computer-Assisted Intervention – MICCAI 2015*, N. Navab, J. Hornegger, W. M. Wells, and A. F. Frangi, Eds., Cham: Springer International Publishing, 2015, pp. 234–241. doi: 10.1007/978-3-319-24574-4_28.

[17] K. He, X. Zhang, S. Ren, and J. Sun, "Deep Residual Learning for Image Recognition," Dec. 10, 2015, *arXiv*: arXiv:1512.03385. doi: 10.48550/arXiv.1512.03385.

[18] A. Kirillov *et al.*, "Segment Anything," Apr. 05, 2023, *arXiv*: arXiv:2304.02643. doi: 10.48550/arXiv.2304.02643.

[19] N. Ravi *et al.*, "SAM 2: Segment Anything in Images and Videos," Aug. 01, 2024, *arXiv*: arXiv:2408.00714. doi: 10.48550/arXiv.2408.00714.

[20] S. J. Pan and Q. Yang, "A Survey on Transfer Learning," *IEEE Transactions on Knowledge and Data Engineering*, vol. 22, no. 10, pp. 1345–1359, Oct. 2010, doi: 10.1109/TKDE.2009.191.

[21] K. Jin *et al.*, "FIVES: A Fundus Image Dataset for Artificial Intelligence based Vessel Segmentation," *Sci Data*, vol. 9, no. 1, p. 475, Aug. 2022, doi: 10.1038/s41597-022-01564-3.

[22] C. Tan, F. Sun, T. Kong, W. Zhang, C. Yang, and C. Liu, "A Survey on Deep Transfer Learning," Aug. 06, 2018, *arXiv*: arXiv:1808.01974. doi: 10.48550/arXiv.1808.01974.

[23] G. Balakrishnan, A. Zhao, M. R. Sabuncu, J. Guttag, and A. V. Dalca, "VoxelMorph: A Learning Framework for Deformable Medical Image Registration," *IEEE Trans. Med. Imaging*, vol. 38, no. 8, pp. 1788–1800, Aug. 2019, doi: 10.1109/TMI.2019.2897538.